\documentstyle[aps,multicol,graphicx,epsfig]{revtex}

\begin{document}
\draft

\title{Materials with Colossal Dielectric Constant: Do They Exist?}

\author{P. Lunkenheimer$^{1}$, V. Bobnar$^{1,2}$, A. V. Pronin$^{1,3}$, A. I.
Ritus$^{3}$, A. A. Volkov$^{3}$, A. Loidl$^{1}$}

\address{$^{1}$Experimentalphysik V, Elektronische Korrelationen
und Magnetismus, Institut f\"{u}r Physik, Universit\"{a}t Augsburg, D-86159
Augsburg, Germany%
\\
$^{2}$Jo\v{z}ef Stefan Institute, P.O. Box 3000, SI-1001 Ljubljana, Slovenia%
\\
$^{3}$ Institute of General Physics, Russian Academy of Sciences, 119991
Moscow, Russia}
\date{}

\maketitle

\begin{abstract}
Experimental evidence is provided that colossal dielectric constants $%
\varepsilon ^{\prime } \geq 1000$, sometimes reported to exist in a broad temperature
range, can often be explained by Maxwell-Wagner type contributions of
depletion layers at the interface between sample and contacts, or at grain
boundaries. We demonstrate this on a variety of different materials. We
speculate that the largest intrinsic dielectric constant observed so far in
non-ferroelectric materials is of order $10^{2}$.
\end{abstract}


\begin{multicols}{2}
\columnseprule0pt \narrowtext

Materials exhibiting a colossal dielectric constant (CDC), $\varepsilon
^{\prime }>10^{3}$, have recently gained considerable attention. CDC
behavior is of technical importance for applications using high-$\varepsilon 
$ electronic materials, such as random access memories based on capacitive
elements. Fundamental interest was initiated by the observation of CDC
behavior in some high-$T_{c}$ parent compounds \cite{htc,mazarra}. Indeed, CDC
behavior may indicate a colossal polarizability, which was invoked in early
polaronic and bipolaronic models as a possible mechanism for high-$T_{c}$
superconductivity \cite{muller}. During the last decade, similar observations of
CDC behavior have been reported in an increasing number of materials, such
as transition metal oxides \cite{oxides,ramirez,homes}.

Large dielectric constants are expected for ferroelectrics in a narrow
temperature range close to $T_{c}$, or for systems with hopping charge
carriers yielding a dielectric constant that diverges towards low
frequencies. However, in various recent reports \cite{htc,mazarra,oxides,ramirez,homes} giant values of the
dielectric constant were claimed to persist over broad temperature ranges
and, when plotted as a function of frequency, revealing an almost constant
low-frequency value and a step-like decrease of the dielectric constant
towards higher frequencies. This step-like decrease, which is accompanied by
a loss peak in the imaginary part of the permittivity, $\varepsilon ^{\prime
\prime }$, shifts exponentially to higher frequencies with increasing
temperature, characteristic of Debye-like dipolar relaxation with a
thermally activated relaxation rate. Several intrinsic physical
interpretations have been given. Examples include almost incipient
ferroelectricity in high-$T_{c}$ materials \cite{mazarra}, highly polarizable
relaxation modes \cite{ramirez}, or a relaxor-like slowing down of dipolar
fluctuations in nano-size domains \cite{homes}.

In the present letter we provide evidence that many of these observations
are not intrinsic in origin and we speculate that most, if not all, of the
CDCs reported so far are based on Maxwell-Wagner type extrinsic effects \cite{mw}. 
We will promote the notion that the most natural explanation
of apparent CDCs are contact effects, and that in ceramic samples grain
boundary effects may play a similar role and further ''enhance'' the
dielectric constant. At these interfaces (metal to insulator contacts,
inter-grain boundaries) depletion layers are formed yielding Maxwell-Wagner
type relaxations when measured by standard dielectric techniques that use
metallic electrodes and two point contact configurations. Thus, while some
of the reports may indeed be based on intrinsic effects, 
extrinsic effects have to be excluded by carefully studying the
materials using different sample geometries, different contact
configurations, and analyzing the results in terms of electronic networks as
considered in detail decades ago \cite{jonscher,ross}.
In the present letter, we propose a simple network that is
derived from the models outlined by the classical works of Jonscher \cite{jonscher}, 
Macdonald \cite{ross}, and others, and also from our years of experience
in dielectric spectroscopy on doped semiconductors. We show examples on a 
series of different materials, where we observed CDCs,
all of which being due to contact effects. Most of the investigated samples are single
crystals, but we also provide results on a ceramic sample to evidence
effects of grain boundaries on the dielectric response. Finally, in addition
to the analysis in terms of electronic networks, we will make some
suggestions on experiments that exclude extrinsic effects.

The inset to Fig. 1a shows the equivalent circuit that describes the main
features of the dielectric response of almost all doped or dirty
semiconductors. The circuit consists of a leaky capacitor connected in
series with the bulk sample. As indicated in the equivalent circuit, the
intrinsic bulk response is given by the sum of dc conductivity $(\sigma
_{dc})$, frequency-dependent ac conductivity - for which we use the
universal dielectric response (UDR) ansatz $\sigma _{ac}^{\prime }=\sigma
_{0}\omega ^{s}$, $s<1$ \cite{jonscher,jonnat} - and the high-frequency
dielectric constant $\varepsilon _{\infty }$ \cite{rem}. The UDR is the most
common approach to take into account hopping conductivity of localized
charge carriers \cite{elliott}. Under these assumptions, the intrinsic complex
conductivity $\sigma _{{\rm intrinsic}}^{\ast }=\sigma _{i}^{\prime
}+i\sigma _{i}^{\prime \prime }$ is given by \cite{jonscher,jonnat}
\begin{equation}
\sigma _{i}^{^{\prime }}=\sigma _{dc}+\sigma _{0}\omega ^{s},   \label{eq1}
\end{equation}
\begin{equation}
\sigma _i^{^{\prime \prime }}=\tan (\frac{s\pi }2)\sigma _0\omega ^s+\omega
\varepsilon _0\varepsilon _\infty ,   \label{eq2}
\end{equation}
\noindent where $\varepsilon _{0}$ is the permittivity of free space. From
the conductivity, the complex dielectric permittivity can be calculated by
$\varepsilon ^{\ast }(\omega )=\varepsilon ^{\prime }(\omega )-i\varepsilon
^{\prime \prime }(\omega )=i\sigma ^{\ast }(\omega )/\omega \varepsilon
_{0}$.
This formalism does not assume any dipolar relaxation phenomenon in the 
compound under investigation. The leaky capacitor in the equivalent circuit
of Fig. 1a represents the most common way to model contributions from the
electrode-sample interface \cite{jonscher,ross}. For semiconducting samples, these
arise mainly due to the formation of Schottky barriers in the region close
to the metallic electrodes. If the electron work function in a metal is
higher than in an electron semiconductor, then in the contact region of the
semiconductor the electron concentration is suppressed, and a depletion
layer appears. This relatively thin layer of small conductivity, acts as a
high capacitance in parallel with a large resistor, connected in series to
the bulk sample. But also an accumulation of defects or deviations from
stoichiometry (e.g. oxygen deficiency) near the sample surface may lead to
such a capacitive surface layer.

\begin{figure}[tbp]
\begin{center}
\includegraphics[clip,width=7.0cm]{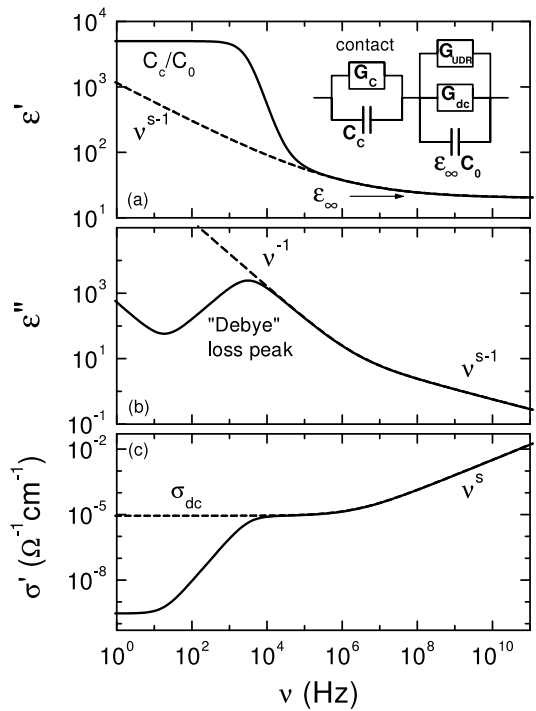}
\end{center}
\caption{Frequency-dependent dielectric response for the equivalent circuit
shown in Fig. 1a. Solid lines: overall response, dashed lines: intrinsic
bulk response as given by eqs. (1) and (2). The circuit parameters have been
chosen to reveal the prototypical behavior of doped semiconductors with
Schottky-barrier type contacts.}
\label{fig1}
\end{figure}

The prototypical dielectric response that results from the equivalent
circuit of Fig. 1a is shown in Fig. 1. The solid lines represent the full
dielectric response, the response of the intrinsic part alone is indicated
by the dashed lines. At low frequencies, $\sigma ^{\prime }(\omega )$ (Fig.
1c) exhibits a step-like increase, which can be ascribed to a successive
bridging of the contact resistance (dominating the low-frequency response)
by the contact capacitance acting like a short at high frequencies. The
intrinsic contribution, following Eq. (1), is revealed only at high
frequencies. Via the relation $\varepsilon ^{\prime \prime }\sim \sigma
^{\prime }/\omega $, the step in $\sigma ^{\prime }(\omega )$ transfers into
a peak in $\varepsilon ^{\prime \prime }(\omega )$ (Fig. 1b), thereby
resembling the response of a Debye-like dipolar relaxation process. It is
accompanied by a large step-like increase of $\varepsilon ^{\prime }(\omega
) $ towards low frequencies (Fig. 1a). At low frequencies $\varepsilon
^{\prime }(\omega )$ is dominated by the very high contact capacitance, $%
C_{c}$, which when divided by the empty capacitance of the sample $C_{0}$,
leads to the apparently colossal values of $\varepsilon^{\prime } =C_{c}/C_{0}>10^{3}$. 
Also for $\varepsilon ^{\prime }(\omega )$, the intrinsic response is
detected only at sufficiently high frequencies. The time constant of the
circuit is approximately $\tau \sim RC_{c}$, with $R$ being the intrinsic
sample resistance. Assuming that the contact capacitance is nearly constant,
the temperature dependence of the time constant is driven by the exponential
increase of the charge carrier density of the semiconducting sample, leading
to $\tau \sim R_{0}C_{c}{\rm exp}(E/T)$, with $E$ being a characteristic
activation energy for charge carriers. Of course the contact step depends on
the thickness and the capacitance of the depletion layer, and is strongly
sample dependent in addition to its temperature dependence. We will show
that in some cases at room temperature the contact step can be shifted well
into the GHz frequency region. Before demonstrating this type of response in
a series of samples, we would like to state that in some cases the
electronic equivalent circuit, shown in the inset to Fig. 1a, is still too
simple \cite{rem}. We have found that in more metallic samples the influence
of inductance has to be taken into account at high frequencies, and
sometimes the contacts are better represented using a distribution of time
constants (several $RC$ values). Furthermore, in ceramic samples other
depletion layers can be formed at the interfaces of grain boundaries,
yielding an additional step in $\varepsilon ^{\prime }(\omega )$ and further
increase of the low-frequency dielectric constant. This will be demonstrated
in one specific sample. However, we provide clear evidence that in most
cases the simple equivalent circuit of Fig. 1a works rather well and
accounts for the CDCs as well as for the relaxational behavior observed in
many semiconducting dielectrics.

Our first example deals with measurements on single crystals of ${\rm {%
CdF_{2}}}$ doped with indium with contacts made from sputtered gold 
(for details, see \cite{sheu,ritus}). The
experimental results and fits performed simultaneously for real and
imaginary part with the simple equivalent circuit of Fig. 1a are shown in
Fig. 2. Here $\varepsilon ^{\prime }(\omega) $ (Fig. 2a) and $\varepsilon ^{\prime \prime}(\omega
) $ (Fig. 2b) are shown in double-logarithmic representation. $%
\varepsilon ^{\prime }(\omega )$ exhibits the characteristic relaxation
steps reaching low-frequency values of about 2000, which are accompanied by
well pronounced loss peaks in $\varepsilon ^{\prime \prime }(\omega )$. For
the higher temperatures, at frequencies below the loss peaks, $\varepsilon
^{\prime \prime }(\omega )$ shows a minimum marking the transition to the
contact-dominated region. Compared to Fig. 1b, this minimum is rather
shallow, but using a distribution of contact barriers a very good agreement
of fits and experimental data could be achieved. For the frequency and
temperature range of Fig. 2, the UDR contribution can be neglected. In this
specific semiconductor we carefully checked the influence of the contacts by
measuring samples of different thickness and employing different contact
electrodes. As an example, in Fig. 2 a second result obtained on the
same sample ($T=126$ K), but with the electrodes separated from the bulk by
a thin insulating layer of mica \cite{sheu,ritus}, is given. Here no Schottky
barriers are formed and the electrode capacitance is simply given by the
mica layer. As shown by the dashed lines, the results can be fitted by the
same equivalent circuit, leading to nearly identical bulk-, but markedly
different contact-parameters. In both cases the intrinsic dielectric
constant $\varepsilon _{\infty }\approx 10$ is far from being colossal or
even unusually large. In order to exclude a Debye dipolar relaxation
process, we also illuminated the sample by laser light \cite{ritus}. It is
obvious, that the Debye relaxation time $\tau _{D}$ should not depend on the
light intensity. In our experiment the ''relaxation frequency'' $\omega
_{p}=1/\tau =1/(RC_{c})$ was proportional to the light intensity due to an
increase of the charge carrier density of the semiconducting sample.

\begin{figure}[tbp]
\begin{center}
\includegraphics[clip,width=7.0cm]{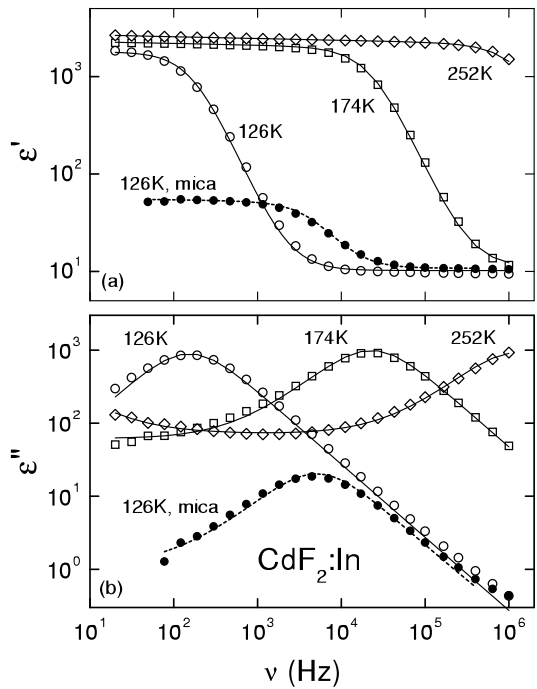}
\end{center}
\caption{Frequency-dependent dielectric response of single-crystalline ${\rm CdF}_{2}$ 
with sputtered gold contacts for three temperatures (open symbols) \protect\cite{ritus}. 
The closed circles show the results at $126~{\rm K}$ for the
same sample with a thin layer of mica between sample and electrodes. The
lines are fits with the equivalent circuit shown in Fig. 1a including a
distribution of contact parameters.}
\label{fig2}
\end{figure}

Figure 3 shows $\varepsilon ^{\prime }(\omega) $ 
for three transition metal oxides, all exhibiting apparently colossal 
dielectric constants at low frequencies. In Fig. 3a, the dielectric response of 
${\rm {LaMnO_{3}}}$ \cite{seeger}, 
the parent compound of all colossal magnetoresistance materials is given. Pure 
${\rm {LaMnO_{3}}}$ is an antiferromagnetic insulator. The finite carrier
density probably results from a slight oxygen excess. The curves in Fig. 3a were taken
at temperatures
below the antiferromagnetic phase transition. Again we find the step-like
decrease of $\varepsilon ^{\prime }$ on increasing frequencies. The solid
lines represent fits using our simple equivalent circuit. Here, the UDR had 
to be included in the fits, leading to a $\omega ^{s-1}$ contribution, 
which smears out the contact-dominated step in 
$\varepsilon ^{\prime }(\omega )$ at high frequencies. For the intrinsic
dielectric constant we obtain, $\varepsilon _{\infty }\approx 15$. For
doped manganates similar results were obtained, yielding somewhat higher,
but certainly not colossal intrinsic dielectric constants \cite{seeger,sichel}.
We also want to refer the reader to our earlier work on single-cystalline  
${\rm {La_{2}CuO_{\text{4+}\delta }}}$, a parent compound of
high-$T_{c}$ materials, which also reveals high non-intrinsic values of  
$\varepsilon ^{\prime }$ \cite{lunkilac}.

Figure 3b shows the results for
single-crystalline ${\rm {SrNbO_{3.41}}}$, which is derived from the high-$%
T_{c}$ ferroelectric compound ${\rm {SrNbO_{3.5}}}$, and has been
characterized as one-dimensional metal  \cite{vidnb}. In this case the
intrinsic dielectric constant is rather high, reaching values of
approximately 100. The low-frequency response approaches 20000, simulating
strong CDC behavior, which, however, arises only from the contacts.

\begin{figure}[tbp]
\begin{center}
\includegraphics[clip,width=9.0cm]{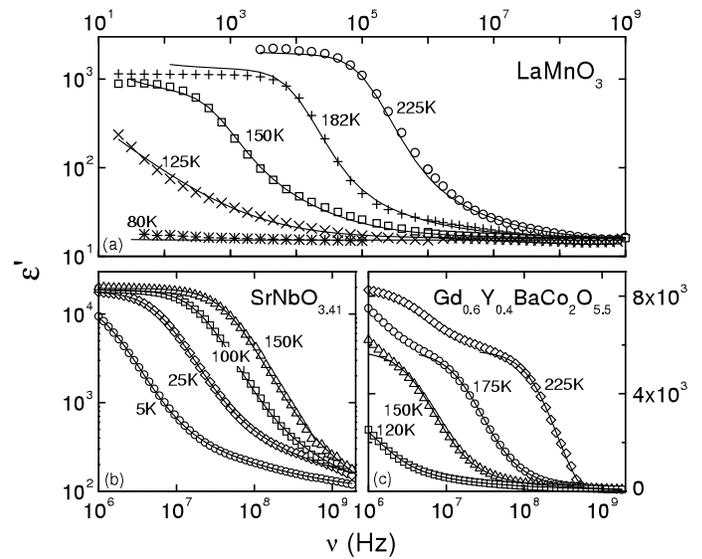}
\end{center}
\caption{Frequency-dependent dielectric constant of three transition-metal oxides 
for various temperatures. The lines are fits with the equivalent circuit shown in 
Fig. 1a. (a) Single-crystalline ${\rm {La_{2}MnO_{3}}}$ \protect\cite{seeger}.
(b) Single-crystalline ${\rm {SrNbO_{3.41}}}$ \protect\cite{vidnb}. (c) Polycrystalline ${\rm {%
Gd_{0.6}Y_{0.4}BaCo_{2}O_{5.5}}}$ \protect\cite{vidcob}. Here the fits were restricted to
the high-frequency step, attributed to grain-boundary contributions \protect\cite{vidcob}.}
\label{fig3}
\end{figure}

Finally, in Fig. 3c we provide experimental evidence that in ceramic samples
the response may be even more complicated. Fig. 3c shows $\varepsilon
^{\prime }(\omega )$ for a perovskite-derived cobaltite \cite{vidcob} revealing
clear evidence for two successive steps. The high-frequency
step, yielding values of $\varepsilon^{\prime }$ close to 6000, is followed
by a further increase towards low frequencies, which elevates $\varepsilon
^{\prime }$ to 8000 at MHz frequencies and $T=225$ K. While the
low-frequency step is due to contact contributions, the second step can be
attributed to grain boundary contributions, which yield a significantly
different effective relaxation rate \cite{vidcob}.

In conclusion, using a variety of different compounds we demonstrated the
occurrence of apparent ''colossal dielectric constants'' at audio and radio
frequencies over a broad temperature range. By an in-depth analysis of the
frequency- and temperature-dependent dielectric response, all these CDCs
were revealed to result from contact or grain boundary effects. In the case
of ${\rm {CdF_{2}:In}}$ we demonstrated how an extrinsic CDC can be excluded
by investigating different contact configurations and different sample
geometries.
We strongly urge that in all further reports of CDC behavior, its intrinsic
nature must be verified by a variation of contact type and/or sample
geometry. Especially if the CDC exhibits a Debye-like relaxation process, an
intrinsic source within the bulk sample is unlikely. In these cases it is
advisable to analyze the data in terms of the equivalent circuits introduced
decades ago \cite{jonscher,ross}, to obtain information on the true dielectric
response of the sample material. This is also important for measurements on
the divergence of the zero-temperature dielectric susceptibility in the
insulating regime, which is one of the most important probes of a
metal-to-insulator transition. To check this behavior experimentally \cite{hess}
it seems important to correct the effective dielectric constants for
contact contributions even at high temperatures.

Here it is also interesting to discuss the early results on charge-density
wave (CDW) systems, which show CDCs at audio- and radiowave frequencies
(see, e.g., \cite{cava}). One is tempted to assume that this behavior also
results from the existence of depletion layers as outlined in this report.
However, a detailed analysis of the dielectric response of ${\rm K}_{0.3}%
{\rm MoO}_{3}$, measured with different types of contacts, was performed by
Cava {\it et al.} \cite{cava}, excluding significant contact contributions.
Furthermore, the unusually large low-frequency response of CDW systems has
been explained theoretically by Littlewood to stem from\ pinned phason
excitations \cite{little}.

In the materials investigated in the present work, the highest intrinsic
dielectric constant observed is $\varepsilon ^{\prime }\approx 100$. What
are the highest values of dielectric constants one can expect in bulk
condensed matter systems? If the origin of the CDCs lies in colossal ionic
or relaxational polarizabilities, they only can be reached in narrow
temperature ranges close to ferroelectric (or relaxor ferroelectric)
transitions, or over broader temperature ranges in incipient ferroelectrics,
like in ${\rm SrTiO}_{3}$ at low temperatures (see, e.g., \cite{viana}). If
the CDCs result from electronic degrees of freedom in semiconductors, within
a nearly free-electron model with an average energy gap $E_{g}$, high values
of $\varepsilon ^{\prime }$\ imply very low values of $E_{g}$  \cite{penn}. In
this case the dc conductivity certainly would dominate the dielectric
response even at moderately elevated temperatures. Thus, with the exception
of CDW systems, we believe that the uppermost limit of an intrinsic
dielectric constant that can occur over broad temperature and frequency
ranges will be of the order of $\varepsilon ^{\prime }\approx 10^{2}$.

This work was supported by the BMBF via VDI/EKM 13N6917-A and partly by
the Deutsche Forschungsgemeinschaft via the SFB 484 (Augsburg).

\end{multicols}

\end{document}